\documentclass[aps,prb,twocolumn,showpacsx]{revtex4}
\usepackage{graphicx}

\newcommand{\be}{\begin{eqnarray}}
\newcommand{\ee}{\end{eqnarray}}

\begin{document}

\title{Photoemission and x-ray absorption study of MgC$_{1-x}$Ni$_3$}
\author{J. H. Kim$^{1}$, J. S. Ahn$^{1}$, Jinsoo Kim$^{2}$, Min-Seok Park$^{3}$, S.
I. Lee$^{3}$, E. J. Choi$^{1,2}$, and S.-J. Oh$^{1,\dag }$}
\address{$^1$School of Physics and Center for Strongly Correlated Materials Research,
Seoul National University, Seoul 151-742, Republic of Korea}
\address{$^2$Department of Physics, University of Seoul, Seoul 130-743, Republic of
Korea}
\address{$^3$National Creative Research Initiative Center for Superconductivity and Department of Physics,
Pohang University of Science and Technology, Pohang 790-784,
Republic of Korea\\}

\begin{abstract}
We investigated electronic structure of MgC$_{1-x}$Ni$_3$ with photoemission
and x-ray absorption spectroscopy. Both results show that overall band
structure is in reasonable agreement with band structure calculations
including the existence of von Hove singularity (vHs)near E$_F$. However, we
find that the sharp vHs peak theoretically predicted near the $E_F$ is
substantially suppressed. As for the Ni core level and absorption spectrum,
there exist the satellites of Ni 2$\mathit{p}$ which have a little larger
energy separation and reduced intensity compared to the case of Ni-metal.
These facts indicate that correlation effects among Ni $3d$ electrons may be
important to understand various physical properties.
\end{abstract}

\maketitle

\address{$^1$School of Physics and Center for Strongly Correlated Materials Research,
Seoul National University, Seoul 151-742, Republic of Korea}
\address{$^2$Department of Physics, University of Seoul, Seoul 130-743, Republic of
Korea}
\address{$^3$National Creative Research Initiative Center for Superconductivity and Department of Physics,
Pohang University of Science and Technology, Pohang 790-784,
Republic of Korea\\}





\begin{section}{Introduction}%
%

Interplay between superconductivity and magnetism is a source of rich solid
state physics. In particular, much attention is being paid to the
superconductors in close vicinity with ferromagnetism (FM). In Sr$_{2}$RuO$%
_{4}$\cite{matz2000}, the p-wave pairing symmetry is thought to be
intimately related with the FM fluctuation of Ru ions. In UGe$_{2}$\cite
{sax2000} and ZrZn$_2$\cite{Pfleiderer2001}, the two phases were found to
coexist. While it was generally believed that FM and SC are mutually
exclusive, these observations suggest that the current understanding needs
to be modified.

Recently, following the discovery of superconductivity in MgB$_{2}$\cite
{naga2001}, intermetallic compound MgC$_{1-x}$Ni$_{3}$ ($x$ = 0.04 $\sim $
0.1) \cite{He2001} was reported to be superconducting at $T_{c}\sim 7K$.
MgCNi$_{3}$ has a cubic antiperovskite structure where C is surrounded by
six Ni ions to form an octahedral cage. Although the material is
paramagnetic at T $>$ T$_c$, Ni ions exhibit ferromagnetic (FM) spin
fluctuation as shown by $^{13}$C NMR experiment\cite{Singer2001}. At T $<$ T$%
_c$, tunneling data\cite{Mao2001} show the zero conductance peak
(ZCP) which may suggest a non-s wave superconducting gap. These
facts seem to indicate that the SC is closely linked with FM in
this system, as also discussed theoretically by Rosner \textit{et
al.}\cite{Rosner2001}. On the other hand, the result from specific
heat experiment\cite{Lin} is consistent with conventional BCS
behavior. Also, Shim \textit{et al.}\cite{Shim2001} showed that
the observed $T_{c}$ is reasonably explained within the BCS
electron-phonon interaction mechanism. Thus, the origin of SC in
MgCNi3 is still controversal.

Studying electronic structure of this material, particularly near the Fermi
energy, is essential to understand the SC. Band calculations\cite
{Dugdale2001,Shim2001,Rosner2001} show that the Ni-3$\mathit{d}$ state is
hybridized with C 2$\mathit{p}$, which dominate density of state (DOS) at E$%
_F$. In particular, it is predicted that a van Hove singularity (vHs) peak
exist very close to the $E_{F}$. The vHs peak gives rise to a large DOS at E$%
_F$ which can be directly related with the superconductive coupling
constant. It is thus important to experimentally probe the vHs peak in
detail. Photoemission spectroscopy (PES) is a powerful tool to investigate
energy-band structures and electron correlation effects. In this work, we
performed PES and x-ray absorption spectroscopy (XAS) measurements of MgCNi$%
_{3}$. We find that overall electronic structure is in general agreement
with band calculation results. The vHs peak is also identified at about 100
meV below E$_F$. However, its intensity is much weaker than the theoretical
predictions. We discuss several possibilities of the suppression and its
implication on the SC.

\end{section}%
%

\begin{section}{Experiment}%
%

Polycrystalline sample used in this experiment was synthesized by the powder
metallurgical technique. Powders were mixed in nominal composition of Mg$%
_{1.2}$C$_{1.45}$Ni$_{3}$. Here, excess Mg and C were added to maximize
carbon incorporation in MgC$_{1-x}$Ni$_{3}$, similar to the previous report.
\cite{He2001} The powders were pelletized, wrapped in a Ta foil, and then
quartz-sealed under vacuum. The sample was reacted for about two hour at 900
$^{\circ }$C and quenched. X-ray diffraction (Rigaku RINT d-max) showed that
the sample is in a single phase. Small amount of MgO impurity and some
unreacted carbon were also identified. Magnetic susceptibility was measured
with a dc SQUID magnetometer (Quantum Design). As shown in Figure 1, the
superconducting onset temperature of this as-grown sample was 6.8 K and the
transition width measured from 10 $\sim $ 90 \% transition was about 0.3 K,
. These results were close to the previous reports of MgC$_{1-x}$Ni$_{3}$.
\cite{He2001} Afterward we put the sample in a high pressure cell and
annealed under 3GPa at 900 $^{\circ }$C for 30 min \cite{Jung2001}to make
densified sample for PES and XAS measurements. After this treatment, the
pellet density increased significantly and was almost identical to its
theoretical value. Figure 1 shows that T$_c$ also increased by about 0.7K
and the transition became sharper.

PES experiments were performed using both He I ($h\nu =21.2$ eV) source at
home laboratory and 120 eV photon at Pohang Light Source (PLS) in Korea. Mg K%
$_{\alpha }$ line ($h\nu =$ 1253.6 eV) was used in x-ray photoemission
spectroscopy (XPS). The resolution for He I, 120 eV and Mg K$_{\alpha }$ are
40 meV, 100 meV and 1eV, respectively. Also, we carried out Ni $L_{3}$-edge
XAS at PLS. Samples were \textit{in-situ} fractured to obtain clean surface.
Base pressure was 2$\times $10$^{-10}$ torr. For the energy calibration, a
bulk palladium was measured at the same time.

\end{section}%
%

\begin{section}{Result and Discussion}%
%

Inset in Figure 2 shows PES data taken with 120 eV photon source. The peak
centered at 1eV below E$_{F}$ corresponds to the Ni 3$\mathit{d}$ derived
conduction band. After the high pressure (HP) treatment, the spectrum below
3eV decreased substantially. In general, photoemission intensity below the
conduction band arises from non-intrinsic sources such as grain boundary or
surface contamination. Decrease of it, together with the T$_{c}$
enhancement, suggests that the sample quality is improved by the HP
sintering. Note, however, that the band structure above 3eV is almost
independent of the sintering. Afterwards, we show the data of the HP
sintered sample. Even the high-pressure sintered sample shows structures
around $\sim $ 6 eV binding energy which we presume mostly come from remnant
MgO precipitates or grain boundaries. It may also be related with the
structural inhomogeneity\cite{Li2002} and the nanoprecipitates\cite{Coo2002}
which were recently reported in this system.

To see the spectra near E$_{F}$ more closely, we took spectrum using He I
source (Figure 2, solid circles). For comparison, we also show predicted
spectra from the three band calculations by Shim \textit{et al.}\cite
{Shim2001} (dotted), by Rosner \textit{et al.} \cite{Rosner2001} (dash dot)
and by Dugdale \textit{et al.}\cite{Dugdale2001} (short dot). To get these
lines, the theoretical densities of states were first convoluted by the
Fermi-Dirac distribution, followed by broadening procedures: the
energy-dependent Lorentzian broadening with $\sim $ $\alpha \left|
E-E_{F}\right| $ ($\alpha =0.3$) and the Gaussian instrumental broadening of
40 meV linewidth.\cite{convol} The curves were normalized to give the same
integrated spectral weight as the experimental data. The short dash-dotted
line shows the background based on the Shirley method.

In the data, four features are observed at $\sim $ 2.7 eV, 1.2 eV, 0.7 eV
and 0.1 eV. The first three features agree roughly with the theoretical
lines, particularly with that by Dugdale\cite{Dugdale2001}. In the curve by
Shim\cite{Shim2001}, 1.2eV peak position is deeper than the data. The curve
by Rosner\cite{Rosner2001} doesn't properly predict the structure centered
at $\sim $ 1.2 eV. According to band calculations, the three features are
from non-bonding Ni 3$\mathit{d}$ states.

The peak located very close to E$_{F}$ (0.1eV) corresponds to the vHs as
predicted in the theoretical lines. Note, however, that the peak height is
much smaller than the predictions. Rough estimation of the peak intensity
shows about 1/2 $\sim$ 1/4 of the theoretical peaks. The vHs arise from the
saddle point near $\Gamma $ in the band structure. The peak strength is then
determined by the band-curvature or the effective mass at this point. Our
observation suggests that the calculations are overestimating the peak
intensity or equivalently underestimating the curvature. Also note that the
height is different significantly among the three theory results, which
shows that band structure near the saddle point depend sensitively on the
calculation methods. Thus, correct estimation of the peak strength seems a
non-trivial work.

In spite of the uncertainty in the calculation, the observed peak is smaller
than any of the three curves. This may suggest that the peak is suppressed
for reasons not accounted for in the band theory.\cite{peak} For example,
when electron-phonon interaction or electron-electron interaction are
present, part of spectral weight will shift to higher energy.  YNi$_{2}$B$%
_{2}$C, a superconductor which bears some similarity with MgCNi$_{3}$ where
Ni-B and Ni-C bondings are  important, is another system that has vHs close
to E$_{F}$. There, the observed peak is also suppressed compared with
theoretical prediction and Kobayashi et al.\cite{Fujimori1996} interpreted
it in terms of electron-electron or electron-phonon interaction.

It is also possible that the spectral weight suppression may be due to
matrix element effects.\cite{matrix,nahm} In this case, the peak intensity
will depend on the photon energy. We thus performed PES using various photon
energy from 40 eV to 150 eV , but the spectra didn't change.\cite{extrinsic}
Surface effect to which UPS is somewhat sensitive is another possible
source. The vHs is a bulk property which arises from the saddle point ($%
\Gamma )$ of the Fermi surface. As one approaches the surface, the band
structure will change and the vHs feature may possibly be smeared.\cite
{private}

Figure 3 shows Ni 2$\mathit{p}$ core-level photoemission spectrum of MgC$%
_{1-x}$Ni$_{3}$. The main peaks, corresponding to Ni 2$\mathit{p}$$_{3/2}$
and 2$\mathit{p}$$_{1/2}$, respectively, are accompanied by the weak
satellites at higher binding energies. The existence of this satellite
structure signals the presence of $\mathit{d}$-$\mathit{d}$ electron
correlation effect, since such satellite structure originates from the
two-hole bound state. For comparison with related compounds, we show Ni 2$%
\mathit{p}$ spectra of Ni-metal\cite{handbook} and YNi$_{2}$B$_{2}$C\cite
{Fujimori1996} together with our data in the inset. Note that in MgC$_{1-x}$%
Ni$_{3}$ the satellite position of Ni 2$\mathit{p}$$_{3/2}$ is a little
large and its intensity is largely reduced compared with Ni-metal. In the
first order approximation, the relative satellite position and the intensity
represent the $\mathit{d}$-$\mathit{d}$ correlation energy and the number of
$\mathit{d}$ holes, respectively. Thus our observations imply that the
correlation energy is a little farther apart and the $d$ hole number is
smaller in MgC$_{1-x}$Ni$_{3}$ compared with Ni-metal.

In MgC$_{1-x}$Ni$_{3}$, Ni is strongly covalent bonded with C and then
charge transfer from the Ni atoms to C will occur. This will result in the
Ni 2$\mathit{p}$ core level shift to higher binding energy. The observed
shift in YNi$_{2}$B$_{2}$C is explained in this manner.\cite{Pellegrin}
However, the binding energy of Ni 2$\mathit{p}$ is almost similar to that of
Ni metal. This seems to suggest that there is a large reduction of the
binding energy due to, probably, a screening of core hole by free carriers.
In fact, the CNi$_{3}$ cage is fully charged by the two electrons donated by
the Mg$^{+2}$ ion and thus Ni hole will be effectively screened.

Figure 4 shows Ni $L_{3}$-edge XAS spectrum of MgC$_{1-x}$Ni$_{3}$. It is
compared with the calculated Ni PDOS above the $E_{F}$ in the band
calculation result by Shim \textit{et al.}.\cite{Shim2001} To get the
theoretical curve (dash-dotted line), the Ni partial density of states are
broadened by convoluting with the similar way as we did in Figure 2. But in
this case we used the Lorenzian broadening with linewidth $\sim $ $\alpha $$%
(E-E_{F})$$^{2}$ ($\alpha =0.2$) and the Gaussian broadening 1.6 $eV$. The
two curves are normalized to give the same integrated spectral weight. We
see that overall structure is in reasonable agreement with the band
calculations. The peak just above E$_F$ is due to the unoccupied part of Ni 3%
$\mathit{d}$ band. But we also observe the satellite structures around $%
\mathit{h\nu}$ = 863 eV in the experiments, which represents the correlation
effects.

\end{section}%
%

\begin{section}{Conclusion}%
%

In conclusion, we have performed XPS, XAS, and UPS measurements to study the
electronic structure of MgC$_{1-x}$Ni$_{3}$. The satellite structure seen in
Ni 2$\mathit{p}$ XPS spectrum suggests that the 3$\mathit{d}$ electron
correlation is substential . L$_{3}$-edge XAS and UPS data show that the
position and width of the Ni 3$\mathit{d}$ derived valence band is in
reasonable agreement with theoretical calculation results. The vHs peak is
found at $\sim $ 0.12 eV below the Fermi energy. Its spectral weight is
largely suppressed compared with theoretical predictions and we suggested
various possibilities of the suppression including the electron-electron and
electron-phonon interaction.

\end{section}%
%

\begin{section}*{Acknowledgments}%
%

We thank H.C. Kim and H.C. Lee for sample characterization at KBSI. \ We
also appreciate K. -J. Kim for the work at PLS. This work was supported by
the KOSEF through the CSCMR.

\end{section}%
%

\begin{figure}[tbph]
\centering
\includegraphics[width=8cm] {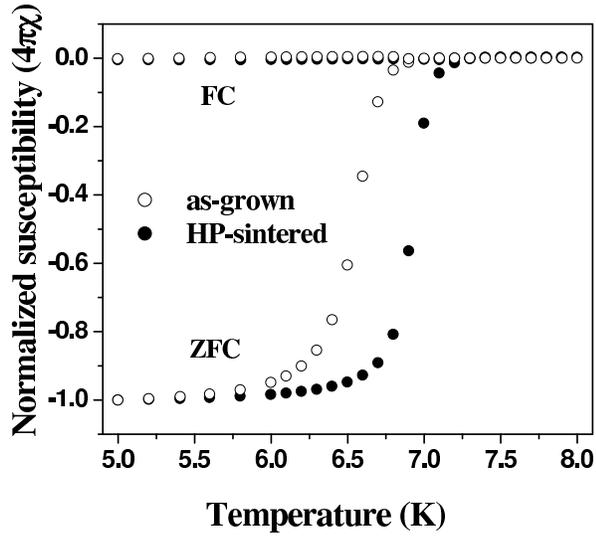}
\caption{Normalized magnetic susceptibility from the measured low-field magnetization $M(T,H=10 Oe)$ of as-grown(solid circles) and sintered at 900$%
^{\circ}$C under 3GPa (open circles).}
\label{Fig:1}
\end{figure}

\begin{figure}[tbph]
\centering
\includegraphics[width=8cm] {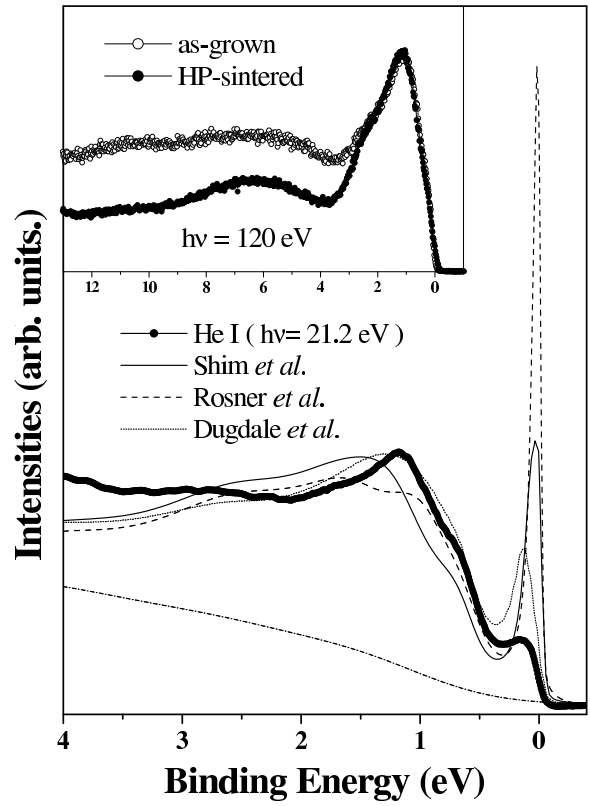}
\caption{ He I UPS spectrum ($\bullet $) of MgC$_{1-x}$Ni$_{3}$ at room
temperature. All lines are theoretical spectra derived from the
band-structure calculations. The short dash-dotted line shows the
background. Inset: Valence region photoemission spectra with photon energy
120 eV of as-grown (solid circles) and sintered at 900$^{\circ}$C under 3GPa
(open circles).}
\label{Fig:2}
\end{figure}

\begin{figure}[tbph]
\centering
\includegraphics[width=8cm] {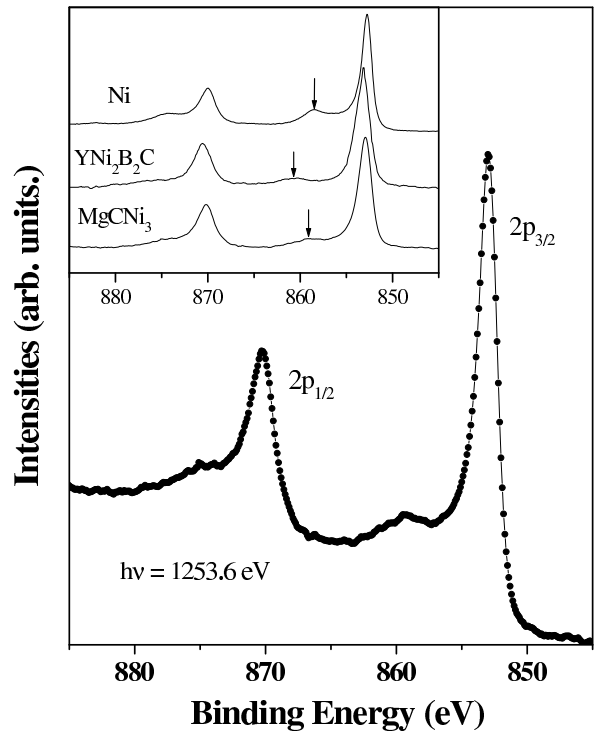}
\caption{Ni $2p$ core-level photoemission spectrum of MgC$_{1-x}$Ni$_{3}$.
Inset: Ni $2p$ spectra of Ni-metal\protect\cite{handbook} and YNi$_{2}$B$%
_{2} $C\protect\cite{Fujimori1996} are compared with that of MgC$_{1-x}$Ni$%
_{3}$. }
\label{Fig:3}
\end{figure}

\begin{figure}[tbph]
\centering
\includegraphics[width=8cm] {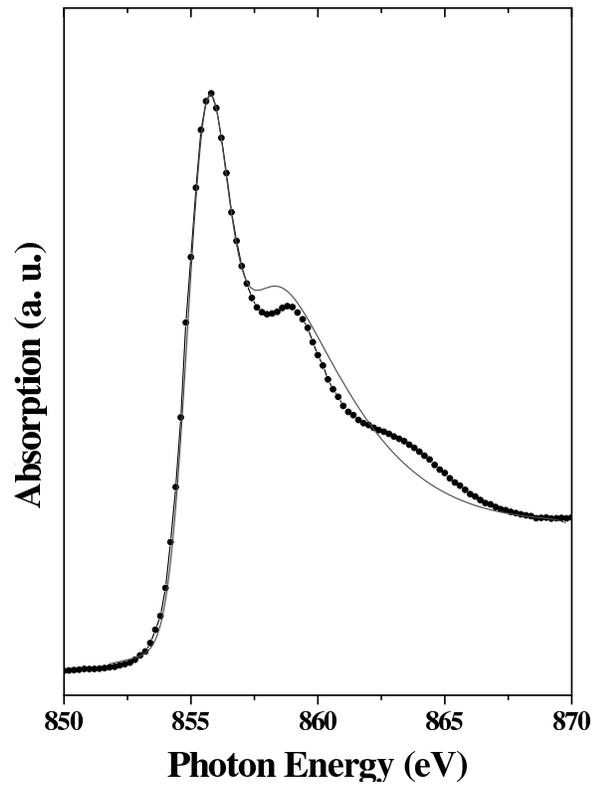}
\caption{Ni $L_{3}$-edge XAS spectrum of MgC$_{1-x}$Ni$_{3}$ ($\bullet $) .
The band calculation result (dash-dotted line) by Shim \textit{et al}.
\protect\cite{Shim2001} is compared. See text for the convolution.}
\label{Fig:4}
\end{figure}

\end{document}